\documentclass[adp,fleqn]{w-art}
\usepackage{times}
\usepackage{w-thm}
\usepackage{amssymb}
\usepackage[]{graphicx}
\chardef\bslash=`\\ 

\begin{document}
\DOIsuffix{theDOIsuffix}
\Volume{XX} \Issue{X} \Copyrightissue{XX} \Month{XX}
\Year{XXXX} 
\pagespan{1}{}
\Receiveddate{1 September 2009} 
\Accepteddate{5 September 2009}



\title[Random energy landscape in organic glasses]{Organic glasses: cluster structure of  the random energy landscape}


\author[S.V. Novikov]{Sergey V. Novikov\footnote{E-mail: {\sf novikov@elchem.ac.ru}, Phone: +7\,495\,952\,2428,
     Fax: +7\,495\,952\,5308}}
\address[]{A.N. Frumkin Institute of Physical Chemistry and
Electrochemistry, Leninsky prosp. 31, 119991 Moscow, Russia}
\begin{abstract}
An appropriate model for the random energy landscape in organic
glasses is a spatially correlated Gaussian field. We calculated
the distribution of the average value of a Gaussian random
field in a finite domain. The results of the calculation
demonstrate a strong dependence of the width of the
distribution on the spatial correlations of the field.
Comparison with the simulation results for the distribution of
the size of the cluster indicates that the distribution of an
average field could serve as a useful tool for the estimation
of the asymptotic behavior of the distribution of the size of
the clusters for "deep" clusters where value of the field on
each site is much greater than the rms disorder. We also
demonstrate significant modification of the properties of
energetic disorder in organic glasses at the vicinity of the
electrode.
\end{abstract}
\maketitle





\section{Introduction}
\label{sect1}

Most disordered organic materials are molecular glasses having
very low concentration of intrinsic free carriers. They
demonstrate measurable conductivity only after injection of
carriers by the action of the laser pulse or strong electric
field. At the same time, organic glasses usually have high
concentration of dipoles and quad\-rupoles. Due to almost zero
concentration of free carriers and lack of screening such
molecules provide long range contribution to the overall
energetic disorder for charge carriers. Long range sources
inevitably lead to the strong spatial correlation of the random
energy landscape $U(\vec{r})$: in organic polar materials
(dipolar glasses) correlation function of random energy
$C(\vec{r}) = \left<U(\vec{r})U(0)\right>$ decays as $1/r$ and
in nonpolar materials (quadrupolar glasses) it decays as
$1/r^3$ \cite{1,Dunlap:542,2}. In this paper the most attention
will be paid to cluster structure of the dipolar glass.	Using
an exact analytic calculation as well as computer simulations
it was shown that in the dipolar glass the random energy
$U(\vec{r})$ is a Gaussian random field if the average distance
between dipoles is not significantly greater than the minimal
distance $a$ (a lattice scale)
\cite{Dieckmann:8136,Novikov:877e}.

Long range correlations mean that sites with close values of
random energy tend to group together and form clusters (cluster
is defined as a set of connected sites, where all of them have
site energy $U$ greater than some boundary energy $U_0$). Until
recently very little has been known about cluster distribution
on size $n_s$ in correlated Gaussian medium (here $s$ is the
number of sites in a cluster). We suggested an approximate
method for the analytic estimation of $n_s$ for "deep" clusters
with $U_0 \gg \sigma$ (here $\sigma$ is rms disorder and, by
definition, $C(0)=\sigma^2$). It turns out that a very simple
calculation of the distribution of the average value of the
correlated Gaussian field in finite domain containing $s$ sites
provides  a very accurate estimation of the leading asymptotic
for the distribution of large clusters with $s \gg 1$ and  $U_0
\gg \sigma$.

We also  discuss briefly how the correlated properties of the
energetic disorder in dipolar glasses are modified in the
vicinity of a conducting electrode.

\section{Distribution of domain averages for Gaussian energy landscapes}
\label{sect2} The simplest domain average is the distribution
$P_{V}(U_{0})$ of the average value $U_0$ of the random energy
$U(\vec{r})$ in a domain with volume $V$ (here we consider a
spatial average and use the same notation $U_0$ for the average
energy)
\begin{equation}
P_{V}(U_{0}) = \left<\delta\left(\frac{1}{V} \int{d\vec{r}U(\vec{r})f_V(\vec{r})} - U_{0}\right)\right>,
\label{def1}
\end{equation}
where $f_V(\vec{r})$ equals $1$ inside the domain and $0$
outside. For the Gaussian random field this distribution can be
calculated in the exact way \cite{6}
\begin{equation}
\label{result}
P_{V}(U_{0}) = \frac{V}{\sqrt{2\pi K}}\exp \left(-\frac{U_0^2 V^2}{2K}\right),\hskip10pt
K = \int{d\vec{r}d\vec{r}_1f_V(\vec{r})C(\vec{r}-\vec{r}_1)f_V(\vec{r}_1)}.
\end{equation}

For the uncorrelated Gaussian field $C(\vec{r})\propto
\delta(\vec{r})$, hence, $K\propto V$ and does not depend on
the shape of the domain. In the opposite case of the very
strong spatial correlation, where $C(\vec{r})\approx
\textrm{const}$ at the microscopic scale, $K\propto V^2$ and
again does not depend on the shape of the domain. This extreme
example looks quite unusual but, nonetheless, this is exactly
the case for the Gaussian random energy landscape provided by
the random uncorrelated distribution of static charges if
concentrations of charges of the opposite sings are equal
\cite{charges}. In that case for the transport layer of the
finite thickness $L$ the only relevant spatial scale for the
correlation function is $L$ and $K\propto V^2$ for $V\ll L^3$.
We can conclude that the whole variety of the dependences of
$K(V)$ is bounded by these two limiting cases. Of course, in
the typical situation $K$ does depend on the shape of the
domain, and these extreme cases are exceptional.

The most studied case (apart from the trivial case of the
uncorrelated Gaussian field) is the dipolar glass model, where
$C(\vec{r})\approx A\sigma^2 a/r$ with $A\approx 0.76$
\cite{1}. In that case for the spherical domain with radius
$R_0$ \cite{6}
\begin{equation}
K = \frac{32\pi^2}{15}Aa \sigma^2 R_0^5\propto V^{5/3}.
\label{kappa_s}
\end{equation}
If a domain has an arbitrary shape but still could be
characterized by a single linear scale $l\propto V^{1/3}$, then
$K \propto V^{5/3}$ just because of dimensionality argument,
though the coefficient of proportionality depends on the actual
shape of the domain.

\begin{figure}[htb]
\begin{minipage}[t]{.45\textwidth}
\includegraphics[width=\textwidth]{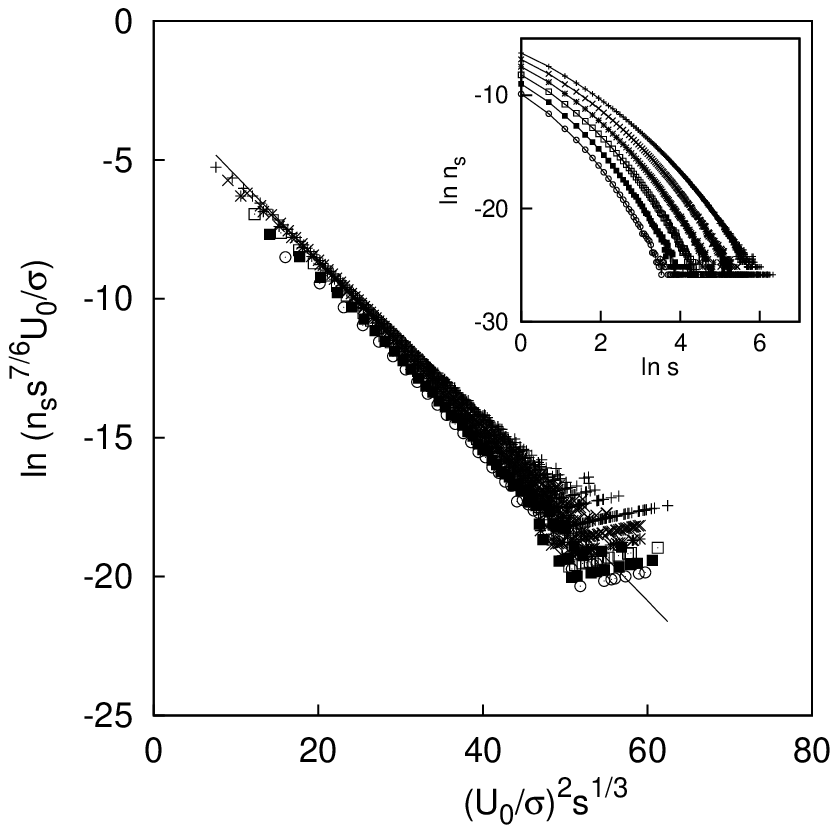}
\caption{Cluster numbers $n_s$ for the dipolar-like
Gaussian field. Threshold energy
$U_0/\sigma$ varies from 2.75 to 4.0 (with the step 0.25)
from the topmost curve downwards (inset).
In proper coordinates
all curves approximately collapse to the uniform straight line
with the slope $-0.31$ in good agreement with (\ref{erfc4}).}
\label{fig5}
\end{minipage}
\hfil
\begin{minipage}[t]{.45\textwidth}
\includegraphics[width=\textwidth]{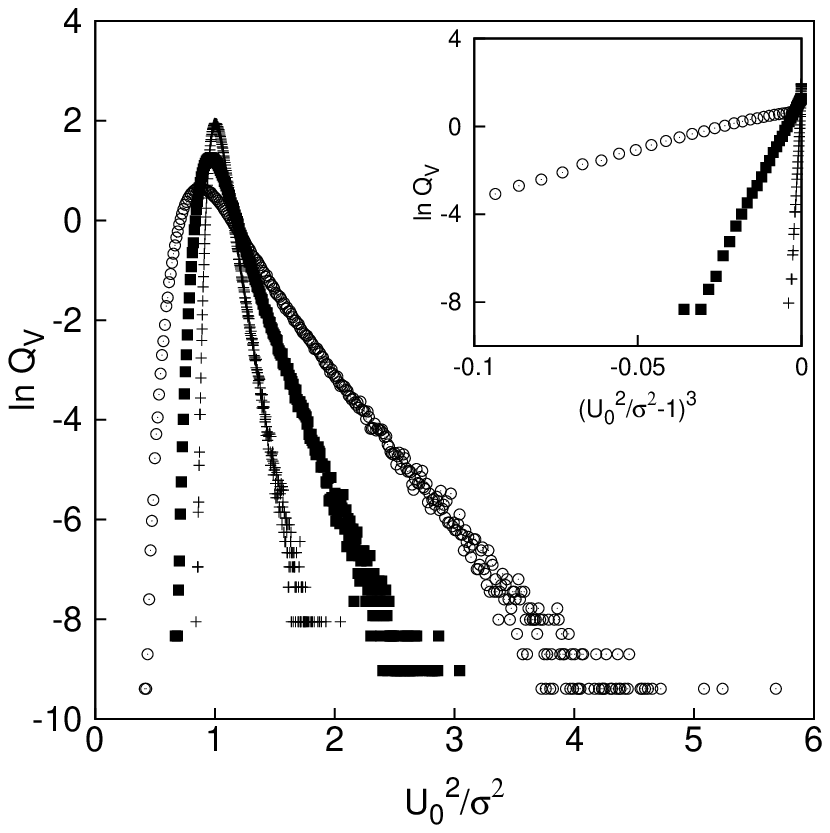}
\caption{Distribution of average squared field $U_0^2$ in the dipolar glass for $R_0$ equals to $5a$, $10a$, and $20a$ (circles,
squares, and crosses, correspondingly). Inset shows the plot of the low $U^2_0$ tails.}
\label{fig6}
\end{minipage}
\end{figure}

Importance of the distribution $P_{V}(U_{0})$ stems from the
fact that it can serve as a good estimation for the number of
true clusters if $U_0\gg\sigma$. Indeed, the number of
spherical domains $n_V(U_0)$ per unit volume, having an average
energy greater than $U_0$, is approximately equal to
\begin{equation}
n_V(U_0) \approx \frac{1}{V}\int_{U_0}^\infty dU P_V(U) =
\frac{1}{2V}\hskip2pt {\rm erfc}\left(\frac{U_0 V}{\sqrt{2K}}\right)
\xrightarrow[]{U_0 \gg \hskip2pt \sigma} \frac{ \sqrt{2K}}{U_0 V^2
\sqrt{\pi}}\hskip2pt \exp\left(-\frac{U_0^2 V^2}{2K}\right),
\label{erfc}
\end{equation}
here the coefficient $1/V$ reflects the number of
non-overlap\-ping independent domains in a finite sample. One
may expect that (\ref{erfc}) gives a reasonable estimation for
the number  $n_s$ of the true clusters, i.e. domains, where
$U(\vec{r}) > U_0$ everywhere (assuming $V=a^3 s$), at least
for the leading term of the asymptotic dependence of $n_s$ on
$s$ (the very use of the continuous model of the random medium
suggests that our consideration is valid only for $s\gg 1$). In
addition, this estimation could be valid only for clusters with
$U_0\gg\sigma$, far away from the percolation threshold,
because we consider the distribution of the average field in
the most compact domain (a sphere). In the vicinity of the
percolation threshold clusters typically have a fractal-like
structure \cite{Stauffer:book}. If this assumption is true,
then for the dipolar-like Gaussian field
\begin{equation}
n_s \propto \frac{\sigma}{U_0s^{7/6}}\exp\left(-B_{d}\frac{U_0^2}{\sigma^2} s^{1/3}\right),
\hskip10pt B_{d}=\frac{5}{4A(36\pi)^{1/3}}=0.34...
\label{erfc4}
\end{equation}
We compared  (\ref{erfc4}) with the simulation data and found
that it provides a good approximation for the true cluster
numbers (see Fig. \ref{fig5}). In addition, this approach gives
a well-known exact asymptotics \cite{5} for the uncorrelated
Gaussian field.

Distribution of the average field in the domain for the
Gaussian random field has a Gaussian form, irrespectively to
the spatial correlation of the Gaussian field. This is not so
for the distribution of the average squared field
\begin{equation}
Q_{V}(U^2_{0}) = \left<\delta\left(\frac{1}{V} \int{d\vec{r}U^2(\vec{r})f_V(\vec{r})} - U^2_{0}\right)\right>.
\label{def1_2}
\end{equation}
For the uncorrelated Gaussian field this distribution, indeed,
has a Gaussian form for $s\gg 1$
\begin{equation}
Q_{V}(U^2_{0}) = \left(\frac{s}{4\pi\sigma^4}\right)^{1/2}\exp\left[-\frac{s(U_0^2-\sigma^2)^2}{4\sigma^4}\right],
\label{def1_2a}
\end{equation}
 while for the
correlated Gaussian field this is not so. For example, for the
dipolar glass the distribution is highly asymmetric (see Fig.
\ref{fig6}). For $U_0^2 > \sigma^2$ the tail of the
distribution approximately follows the law $\ln Q_V\propto
-(U_0^2-\sigma^2)$, while for $U_0^2 < \sigma^2$ the
corresponding law is $\ln Q_V\propto (U_0^2-\sigma^2)^3$. This
difference directly follows from the strong spatial correlation
of $U(\vec{r})$: inner parts of clusters provide contribution
to $Q_V$ for $U_0^2>\sigma^2$, while the boundaries between
clusters mostly contribute to the tail $U_0^2<\sigma^2$.

\section{Energetic disorder at the interface: how it differs from the bulk}

Electrostatic nature of the energetic disorder in organic
materials has a very nontrivial consequence. Statistical
properties of such disorder at the interface with other
material may change dramatically without any variation of the
spatial or chemical structure of the organic material.

Let us consider the interface of organic glass with the metal
electrode. The electrostatic energetic disorder is directly
proportional to the disorder in the spatial distribution of
electrostatic potential $U(\vec{r})=e\varphi(\vec{r})$,
generated by randomly situated and oriented dipoles or
quadrupoles. In organic
 layers sandwiched  between
 conducting  electrodes distribution of $\varphi(\vec{r})$ must obey a boundary condition at the
electrode surface: at this surface the potential should be a
constant. Thus, at the electrode surface there is no energetic
disorder at all, irrespectively to how disordered is the
material in the bulk. This means that the magnitude of the
dipolar or quadrupolar disorder increases while going away from
the interface, asymptotically reaching its bulk value
$\sigma_b$. Magnitude of the dipolar disorder at the interface
is decreased as (here $z$ is the distance from the interface
plane) \cite{prb}
\begin{equation}\label{sigma2(z)_3}
\sigma^2(z)\approx\sigma^2_b
\left[1-\frac{a_0}{2z}\left(1-e^{-2z/a_0}\right)\right],
\hskip10pt a_0=Aa.
\end{equation}
Asymptotics of the correlation function
$C(z_1,z_2,\vec{\rho}_1-\vec{\rho}_2)=\left<U(z_1\vec{\rho}_1)U(z_2,\vec{\rho}_2)\right>$
for $|\vec{r}_1-\vec{r}_2|\gg a$ can be calculated using the
approach, suggested in \cite{pss}
\begin{equation}\label{z=z'}
    C(z_1,z_2,\vec{\rho})=\sigma^2_b a_0\left(\frac{1}{\sqrt{\rho^2+(z_1-z_2)^2}}-\frac{1}{\sqrt{\rho^2+(z_1+z_2)^2}}\right).
\end{equation}
Here $\vec{\rho}$ is a 2D vector oriented along the interface
plane. Hence, at the interface the dipolar glass is much less
correlated in comparison with the bulk:
$C(z_1,z_2,\vec{\rho})\propto z_1z_2/\rho^3$ for $\rho \gg
z_1,z_2$, while in the bulk $C(r)\propto 1/r$, and clusters are
elongated perpendicular to the interface.

Parameter $K$ for a sphere, located at the distance $z$ from
the interface, can be estimated by a direct generalization of
(\ref{result}) to the case of spatially non-uniform disorder
and has a form $K_\textrm{bulk}-\delta K$ with
\begin{equation}\label{dK}
\nonumber
\delta K=16\pi^2\sigma_b^2 a_0 R_0^5\int_0^\infty\frac{dk}{k^2}\exp\left(-2\frac{kz}{R_0}\right)g^2(k), \hskip4pt
 g(k)=\int_0^1 dt \sqrt{1-t^2}J_1\left(k\sqrt{1-t^2}\right)\cosh(kt),
\end{equation}
\[
\delta K\approx \sigma_b^2 a_0 \frac{V^2}{2z}, \hskip10pt R_0\ll z.
\]
Here $J_1(x)$ is a Bessel function. Magnitude of $K$ at the
interface is smaller than in the bulk, reflecting the decrease
of the spatial correlation.

\section{Conclusion}
In this paper we discussed the cluster properties of the
electrostatic energetic disorder in organic glasses with the
emphasis on the properties of polar materials built by
molecules having permanent dipole moments. We found that for
this disorder the distribution of the average energy could
serve as a good estimation for the true cluster numbers per
lattice site  for large "deep" clusters, where $s\gg 1$ and the
threshold energy $U_0$ is significantly greater than the rms
disorder $\sigma$. Comparison of the analytical results for
$P_V(U_0)$, calculated for a spherical domain, and computer
simulation data for $n_s$ supports the hypothesis that
$P_V(U_0)$ provides the exact leading asymptotic term for
$n_s$. We have also shown that the electrostatic energetic
disorder in organic semiconductors changes dramatically in the
neighborhood of a metal electrode. This phenomenon is of
principal importance for the description of the charge
injection in organic devices.

\begin{acknowledgement}
This work was supported by the ISTC grant 3718 and RFBR grant
 08-03-00125.
\end{acknowledgement}

\end{document}